# Review of Knowledge Management Systems
# As Socio-Technical System

Setiawan Assegaff[1], Ab Razak Che Hussin[2]

[1] Program Pasca Sarjana Magister Sistem Informasi, STIKOM Dinamika Bangsa
Jambi, Indonesia

[2] Faculty of Computer Science and Information System, Universiti Teknologi Malaysia
Skudai, Johor, Malaysia

**Abstract**

Knowledge Management Systems as socio-technical system perspectives has recognized for decades. Practitioners and scholars belief Knowledge Management is best carried out throught the optimization both technological and social-aspect. Lacking of understand and consider both aspects could lead organizations in misinterpretation while developing and implementing Knowledge Management System. There is a need for practical guidance how Knowledge Management System should implement in organizations. We propose a framework that could use by practitioner and manager as guidance in developing and implementing Knowledge Management System as Socio-Technical Systems. The framework developed base on Pan and Scarborough view of Knowledge Management as Socio-Technical system. Our framework consists of: Infrastructure (technology), Info structure (organizational structure) and Info culture (organizational culture). This concept would lead practitioners get clear understand aspect contribute to Knowledge Management System success as Socio-Technical System.

## 1. Introduction

Knowledge recognized as the only asset could contribute in sustainable competitive advantages for organization [1, 2]. Organizations realize to get benefit from knowledge, they should consider in managing organizational knowledge effectively. How organization manage their knowledge known as Knowledge management (KM) [3]. Today's, there are some perspectives in term how KM should be implemented in organizations. The perspectives derive from a different point of view in understanding what the KM is. First perspective is understood the KM more from the perspective of information systems, databases, and knowledge structures, and believe that knowledge is developed and managed according to universal and standardized rules. Social or organizational culture and other social factors are removed from the equation or disregarded outright [4]. Second perspective focus on the flow of information among self-managing groups within an organization that considers a team to be the primary holder of information [5]. The third perspective has a different view of KM, they belief knowledge management as the interaction between physical re-sources, conceptual resources, and social and organizational processes [6].

The implication of different perspective in understands KM leads scholars and practitioners develop and implement different strategies in KM initiatives in organizations. Some of them who beliefs KM from information system and belief knowledge is able to manage focus on Information Technology (IT) in implementation KM [7]. Others scholars and practitioners that belief KM as social and organizational process develop KM with focus on social and cultural aspects [8].

However experiences from different success studies in KM suggest IT-based approaches to KM need to be complemented by social methods undertaking. Research in KM success notes IT alone or social system alone has failed to deliver KM mission in many organizations [9]. This is because both IT and social-system has their own different contribution in delivering KM success. IT is very good delivering explicit knowledge could manage and distribute information effectively and efficiently. The social-system is very good deliver tacit knowledge and support creation of knowledge through social interaction. Both of IT and social system contribution in KM are an equally important. It became the reason why technologies and social system are equally important in KM [10].

Studies in KM as socio-technical system have been done by researchers for decades. Most of studies note the importance of interplay of knowledge management process and organizational context. However, the studies presented the whole spectrum of elements that need to be designed and encouraged in order to create an effective knowledge management system in the organization is still limited [11]. Our studies conduct to contribute to the gap. We propose a framework that could use by scholars and practitioners to aware on critical elements for success KMS implementation from socio-technical perspective.

## 2. Knowledge Management

Organizations around the world became more aware about knowledge. Knowledge is believed as potential asset that could bring sustainable competitive advantages for organization. To gain value from the knowledge, organization should able to manage it effectively. Approach implemented by organizations to manage knowledge recognise as KM. [12]. Todays KM notes as vital integral in business functions. According to Alavi and Leidner [13] KM process consist of four processes, first knowledge creation, knowledge storage/ retrieval, knowledge transfer and knowledge application. Four process in KM will describe below:

- **Knowledge Creation-** knowledge creation related to developing new knowledge or replaces existing knowledge in terms of tacit and explicit knowledge.
- **Knowledge Storage/ Retrieval-** Knowledge storage/ Retrieval include activities such as knowledge residing in various component forms, knowledge structure, codifying the knowledge and store of knowledge to organizational memory.
- **Knowledge Transfer** – Knowledge transfers exist between individual, individual to groups, groups to groups, groups in organizations and across.
- **Knowledge Application**- Knowledge application is an integration of knowledge to organization process or activities such as directives, organizational routines, and self-contained task teams.

IT has been used for a long time in support business process in organizations. IT use in the organization to make numerous contribution such as reducing time, cost, support better services for customers. The practitioner also considers IT to support KM. IT use in KM in various ways. Many applications have developed and use to support KM. Social network software, video/tele-conference, organization directories, e-mail, e-learning, repositories were potential tools in support KM. IT found very potential in support KM. The main function of IT in KM is to support and enabler KM process. Figure bellow describes how IT contributes in KM process.

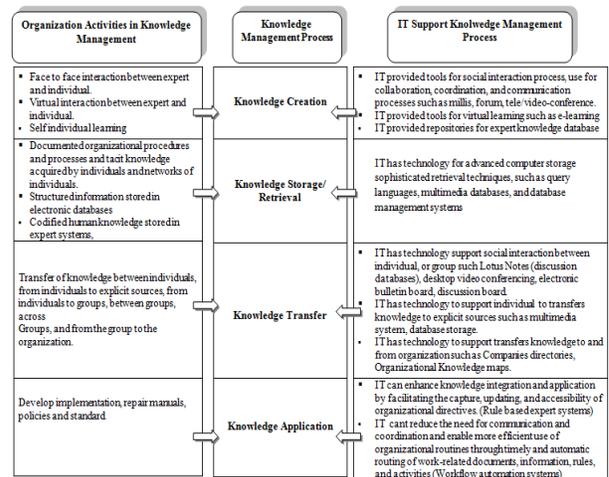

Fig. 1 IT Function support KM Process
(Adapted from Alavi and Leidner, 2001).

Figure 1 above explain how the relation between organization activities, KM process and IT in enabling KM implementation. Each of the social activities of people in the organization in the KM process could support by specific application developers in the organization. Example knowledge creation occurs in social interaction between individual and expert, IT can support this interaction by providing tools like video conference. However IT is not the only aspect that should focus in KM. Social system related to culture and organizational structure are important aspects that managers should pay attention.

## 3. Knowledge Management System

In common, Knowledge Management Systems (KMS) are IT that enables organizations to manage effective and efficient knowledge. Some definition of KMS has been proposed by some researchers. In this study we use the KMS definition by Alavi and Leidner [13]. They defined KMS as a class of information systems applied for managing organizational knowledge. In general KMS would not have differences from other information systems, instead of content and activities by users. KMS would consist of hardware, software, people, and organization environment around it.

To well understand about KMS, it's better for us familiar with the characteristic of KMS. Maier and Hädrich [14] propose a characteristic of KMS, consist of goal, processes, comprehensive platform, advance knowledge, knowledge services, Knowledge instruments, specific knowledge, and participants.

Table 1: Characteristics of KMS
(Maier and Hädrich, 2006)

| KMS Characteristics Component | Explanation of Component |
|---|---|
| Goals | <ul><li>Bring knowledge from the past to bear on present activities, thus resulting in increase levels of organizational effectiveness (Lewis and Minton (1998); Stein and Zwass (1995).</li><li>As the technological part of KM initiative that also comprises person-oriented and organizational instruments targeted at improving the productivity of knowledge work (Maier (2004))</li></ul> |
| Processes | <ul><li>Developed to support and enhance knowledge-intensive task, processes, or projects (Detlor, 2002); Jennex and Olfmann (2003))</li><li>Supported knowledge processes such as, knowledge creation, organization, storage, retrieval, transfer, refinement and packaging, (re)use, revision, and feedback, also called the knowledge life cycle, ultimately to support knowledge work (Davenport et al. (1996))</li></ul> |
| Comprehensive Platform | <ul><li>KMS is not an application system targeted at single KM initiative, but a platform that can be used either as IT to support knowledge processes or integrating base system and repository in which KM application systems are built (Maier (2006))</li><li>There are two platform categories, the first user-centric approach with focus on processes, and IT-centric approach which focuses on base system to capture and distribute knowledge (Jennex and Olfman (2003))</li></ul> |
| Advanced Knowledge Services | KMS are ICT platform consist of a number of integrated services<ul><li>Basic services such as, collaboration, workflow management, document and content management, visualization, search and retrieval (Seifried and Eppler (2000))</li><li>Advanced services such as, personalization, text analysis, clustering and categorization to increase the relevance of retrieved and push information, advanced graphical techniques for navigation, awareness services, shared workspace, and learning services as well as the integration of reasoning about various sauces on the basis of shared ontology ( Bair (1998); Borgoff and Parechi (1998); Maier (2004))</li></ul> |
| Knowledge Instruments | <ul><li>KMS are applied in a large number application area (Tsui, 2003)</li><li>KMS specially support KM instruments (Alavi and Leidner (2001); McDermott (1999); Tsui (2003))</li><li>KMS offers targeted combination and integration of knowledge services that together foster one or more KM instruments (Maier, 2006))</li></ul> |
| Specifics of Knowledge | KMS help to assimilate access to sources of knowledge, and with the help of shared context, increase the breadth of knowledge sharing between persons rather than storing knowledge itself (Alavi and Leidner (2001)) |
| Participants | Users play roles of active, involved participants in the knowledge network forested by KMS (Maier, 2006)) |

## 4. Knowledge Perspective and Its Implication to KM and KMS

There are different views of knowledge lead to different perspectives on how knowledge to be manages. Alavi and Leidner [13] proposed three different views of knowledge, as object, process and capabilities. First knowledge as an object, it related to information access, the implication is the key of KM develops on building and managing information stock/information. If knowledge as process, it means KM should focus on how knowledge/information could be created, share, and distribute among employee in organization. If knowledge is capabilities, KM will lead employee to build their competencies, skill, and produce intellectual capital.

These different views of knowledge have implication on how KMS to design. It brings us to consider three different views to be included in KMS mission. KMS should focus on knowledge as well as focus on people. KMS should have function/feature not only for managing knowledge/information but also to facilitate people to stay in touch, connect together, so they able to share and thinking together among communities [15]. KMS in another world should develop with considering KM as socio-technical system.

Knowledge as an object is very relevant with concept of IT function, as Bath [3] argues that IT can handle data and information efficiently in KM, but IT poor at interpreting information to be knowledge. IT should connect together with people so the social system that needed to create and share knowledge happen. IT can support social-system by provides tools for interaction among member of the communities.

Table 2: Knowledge Perspective and its Implication to KMS
(Alavi and Leidner, 2001)

| Perspectives | | Implications for Knowledge Management (KM) | Implications for Knowledge Management Systems (KMS) |
|---|---|---|---|
| Knowledge vis-à-vis data and information | Data is facts, raw numbers. Information is processed/interpreted data. Knowledge is personalized information. | KM focuses on exposing individuals to potentially useful information and facilitating assimilation of information | KMS will not appear radically different from existing IS, but will be extended toward helping in user assimilation of information |
| State of mind | Knowledge is the state of knowing and understanding. | KM involves enhancing individual's learning and understanding through provision of information | Role of IT is to provide access to sources of knowledge rather than knowledge itself |
| Object | Knowledge is an object to be stored and manipulated. | Key KM issue is building and managing knowledge stocks | Role of IT involves gathering, storing, and transferring knowledge |
| Process | Knowledge is a process of applying expertise. | KM focus is on knowledge flows and the process of creation, sharing, and distributing knowledge | Role of IT is to provide link among sources of knowledge to create wider breadth and depth of knowledge flows |
| Access to information | Knowledge is a condition of access to information. | KM focus is organized access to and retrieval of content | Role of IT is to provide effective search and retrieval mechanisms for locating relevant information |
| Capability | Knowledge is the potential to influence action. | KM is about building core competencies and understanding strategic know-how | Role of IT is to enhance intellectual capital by supporting development of individual and organizational competencies |

## 5. Discussion

### 5.1 KMS as Social-Technical System

The term "socio-technical" was initially introduced by Trist [16] to emphasize the interrelationship between social factors and technological factors in understanding an organization. In organization the Socio Technical System (STS), an organization consists of people and technology, according to the theory, the two systems, technical and social need to work mutually to produce optimized output. After that Pan and Scarborough [2] further developed the concept of socio-technical in KM.

We use the concept and adopt it to KMS as view at figure bellows. KMS as socially constructed, shaped by the between technological and social factors an organizational context. KMS from socio-technical perspectives have two aspects. First is technical aspects, it's about IT (hardware and software use to support KM) or some scholars name it as infrastructure. Second aspects related to social - systems that consist of organizational structure (infrastructure) and Organizational culture (info culture). Detail explanation about infrastructure, infrastructure and info culture will be described in the following section.

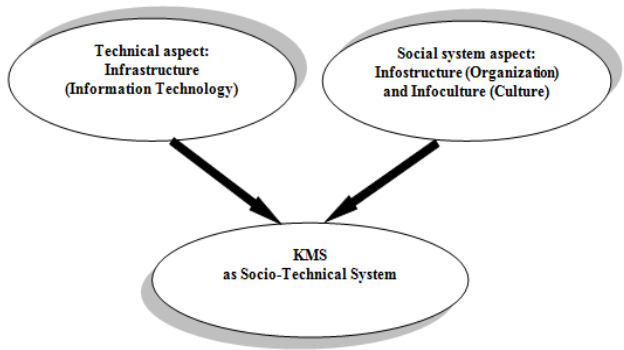

Fig. 2 KM as Socio-Technical System

KMS under socio-technical system perspectives should placed three layers of interaction in implementation. Each of layers influences each others. KM would not success without contribution from that layer [9]. The layers are: (1) infrastructure: hardware/software that enables communications between nodes or members of the network, (2) infrastructure: formal rules governing the exchange between actors in the network through metaphors and common language, and (3) info culture: background knowledge, embedded in social relations surrounding work group processes, that defines the cultural constraints on knowledge and information sharing. We use Pan and Scarborough layer interaction to develop our propose framework of KMS as socio-technical system as described in figure 3.

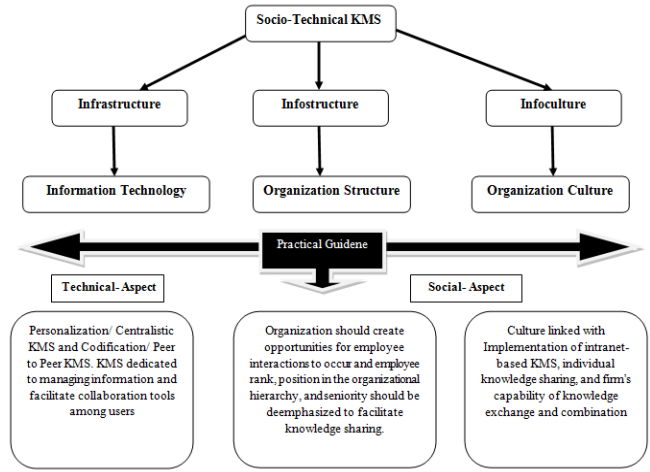

Fig. 3 infrastructure, Infostructure, infoculture as part of Socio-Technical KMS

We discuss more detail the three layers and the implication of them to organizations. To adopt KMS as socio-technical system organization should pay attention to the following discussion of previous three layers.

**Infrastructure layer-** Infrastructure in this concept is defined as the using of IT in support KM implementation. IT is consist of hardware and software. IT uses in KM or KMS have two main functions. The first function is to manage information and the second is to facilitate collaboration among user. Table 3 describes the role of KMS in managing information and facilitates community collaboration.

Table 3: KMS Function

| Managing Information | Facilitating Communities |
|---|---|
| - KMS help user assimilation of information<br>- KMS has a function as gathering, storing, and transferring information/knowledge<br>- KMS provides effective search and retrieval mechanisms for information | - KMS provides access to sources of knowledge rather than knowledge itself<br>- KMS should provide links among sources of knowledge to create wide breadth and deep knowledge flows<br>- KMS enhances intellectual capital by supporting development of individual and organizational competencies |

Applying KMS on managing information known as codification strategy. KMS would have a feature in assimilating information for users. KMS also provides some tools for gathering, storing, and transferring information to and from the system around communities [14,22]. Others advanced functions of KMS is supporting collaboration in teams and communities. KMS would provide tools that link knowledge contributor and seeker, and provide e-learning functionality integrated in KMS. This trend has equal spirit with KMS as socio-technological system philosophy. However Implement purely KMS from technical aspect would not achieve KMS mission without consideration social aspect of KMS [17].

Organizations should consider info structure and Info culture that will describe bellows.

- **Info structure layer-**In previous studies about culture in KM implementation suggest that organizations should create opportunities for employee interactions to occur and employee rank, position in the organizational hierarchy, and seniority should be deemphasize to facilitate knowledge sharing. [18, 19, 23]. Many organization traps in the procedural and structural condition in term of relation between staff and managers. Sometimes the condition would lead employee in a hard position to active communication each others. Without intensive interaction among member in organization knowledge sharing could not exist and deliver KM mission. Only organizations that able break to unsupported environment and develop a conducive environment could deliver KM goal for their organizations.
- **Info culture layer-**Previous studies found that the benefits of new technology infrastructure were limited if long-standing organization values and practices were not supportive of a knowledge sharing process. A culture that emphasized trust has been found to help alleviate the negative effect of perceived of cost of knowledge sharing. Culture linked with Implementation of intranet-based KMS, individual knowledge sharing, and firm's capability of knowledge exchange and combination. Organization with cultures emphasizing innovation more likely to implement intranet KMS [10, 20]. Many scholars believe that culture values creativity continues improvement and the sharing ideas are necessary for KM implementation [23]. From previous researchers, scholars found that management in organizations can manipulate the organization environment by intervention them. They believed Intervention is best approaches way of management to support KM implementation and achieve KM mission. One scholar that proposes the concept is Cabrera and Cabrera [21]. They argue that three potential solution organizational interventions that may increase the employee participation in KS programmers. First restructuring the payoff function, increasing the perceived efficacy of individual contribution, and establishing group identity and promoting personal responsibility. Organizations can restructure the payoff function through decrease the perceive cost of contributing, if the cost is lower than benefit, the employee will motivate to share. Organizations can provide enough time for their employee to explore or contribute their knowledge. Develop user friendly information system and give easy and enough access to knowledge recourses also believed encourages employees to share their knowledge. Organizations also can develop a reward system for motivated employee for knowledge sharing, rewards can be as financial and non financial.

## 6. Conclusion

The Socio - technical system is one of potential perspectives that could consider by practitioner and manager as philosophy in developing and implement KMS success. In this perspectives manager and practitioner should consider three components. Using our framework could bring practitioner get well understand what factors related to three components of socio-technical system of KMS. The first component is infrastructure, infrastructure related to IT. IT in KM has function of managing knowledge and facilitate collaboration among users. Second is infostructure, where an organization should develop a flat structure to enable employee got rich communication channel each others. The last is about infoculture, this is related to an organization's ability to create appropriate environment to encourage employee doing knowledge sharing.

**Setiawan Assegaff**. He graduated from the Gunadarma University, Indonesia, and now as PhD student at the Universiti Teknologi Malaysia, Malaysia. He has experience as lectures in STIKOM Dinamika Bangsa, Jambi, Indonesia for eight years. His special fields of interest included Knowledge Management, Technology Adoption and Computer and Society.

**Ab Razak Che Hussin.** He graduated from the University of Manchester, UK in 2006 after completed his PhD in Trust in e-Commerce. He is now senior lecturer at Universiti Teknologi Malaysia, Malaysia. His special fields of interest included Information System, Web Application, and Trust and Privacy in e-Commerce.